\documentclass[prl,twocolumn, amsmath, amssymb, superscriptaddress]{revtex4-1}
\pdfoutput=1
\usepackage{hyperref}
\usepackage{graphics}
\usepackage[pdftex]{graphicx}
\usepackage{textcomp}

\begin{document}
\title{A nuclear quantum effect with pure anharmonicity and the anomalous thermal expansion of silicon}
\author{D.S. Kim}
\email[Electronic Address: ]{dennis.s.kim@icloud.edu}
\affiliation{California Institute of Technology, Department of Applied Physics and Materials Science, Pasadena, California 91125, USA}
\author{O. Hellman}
\affiliation{California Institute of Technology, Department of Applied Physics and Materials Science, Pasadena, California 91125, USA}
\author{J. Herriman}
\affiliation{California Institute of Technology, Department of Applied Physics and Materials Science, Pasadena, California 91125, USA}
\author{H.L. Smith}
\affiliation{California Institute of Technology, Department of Applied Physics and Materials Science, Pasadena, California 91125, USA}
\author{J.Y.Y. Lin}
\affiliation{Neutron Data Analysis and Visualization Division, Oak Ridge National Laboratory, Oak Ridge, Tennessee 37831, USA}
\author{N. Shulumba}
\affiliation{California Institute of Technology, Department of Mechanical and Civil Engineering, Pasadena, California 91125, USA}
\author{J.L. Niedziela}
\affiliation{Instrument and Source Division, Oak Ridge National Laboratory, Oak Ridge, Tennessee 37831, USA}
\author{C.W. Li}
\affiliation{University of California Riverside, Department of Mechanical Engineering, Riverside, California 92521, USA}
\author{D.L. Abernathy}
\affiliation{Quantum Condensed Matter Division, Oak Ridge National Laboratory, Oak Ridge, Tennessee 37831, USA}
\author{B. Fultz}
\email[Electronic Address: ]{btf@caltech.edu}
\affiliation{California Institute of Technology, Department of Applied Physics and Materials Science, Pasadena, California 91125, USA}
\begin{abstract}
Despite the widespread use of silicon in modern technology, its peculiar thermal expansion is not well-understood.
Adapting harmonic phonons to the specific volume at temperature, the quasiharmonic approximation, has become accepted for simulating the thermal expansion, but has given ambiguous interpretations for microscopic mechanisms.
To test atomistic mechanisms, we performed inelastic neutron scattering experiments from 100--1500\,K on a single-crystal of silicon to measure the changes in phonon frequencies.
Our state-of-the-art \textit{ab initio} calculations, which fully account for phonon anharmonicity and nuclear quantum effects, reproduced the measured shifts of individual phonons with temperature, whereas quasiharmonic shifts were mostly of the wrong sign.
Surprisingly, the accepted quasiharmonic model was found to predict the thermal expansion owing to a large cancellation of contributions from individual phonons.
\end{abstract}
\date{\today}
\maketitle

A quantized harmonic oscillator was Einstein's seminal idea for understanding atom vibrations in solids.
Better accuracy for crystalline solids is achieved when the vibrations are resolved into normal modes.
Each normal mode is quantized, with a zero-point energy and excitations called phonons.
However, harmonic models are limited to quadratic terms in the interatomic potential, and it is well-known that higher order terms are necessary to describe properties of real solids such as thermal conductivity and thermal expansivity.
Despite this knowledge, the necessary and sufficient contributions to non-harmonic effects remain less clear.
A popular approach is the quasiharmonic model (QH), which assumes harmonic oscillators,
but with frequencies renormalized to account for the thermal expansion.
In a quasiharmonic model, the energy of the phonon mode $i$ changes with crystal volume, $V$.
Changes to phonon energies are usually described by a mode Gr\"uneisen parameter, $\gamma_i= - (V\,\partial\varepsilon_i) / (\varepsilon_i \,\partial V)$, where $\varepsilon_i = \hbar \omega_i$ is the phonon energy (and $\omega_i /2\pi$ is the frequency).
A positive $\gamma$ gives a reduction in mode energy with thermal expansion, increasing the vibrational entropy $\Delta S_{\rm vib}$.
At finite temperature, the extra elastic energy from thermal expansion, $\Delta E_{\rm el}$, is offset
by the term $-T \Delta S_{\rm vib}$ in the free energy $\Delta F = \Delta E_{\rm el} - T\Delta S_{\rm vib}$  \cite{BrentFultz:2010ew,Fultz:2014jk}.
For positive $\gamma$, $\Delta F$ is minimized with a positive thermal expansion; for negative $\gamma$, a negative thermal expansion is expected.

\begin{figure*}
\includegraphics[width=\textwidth]{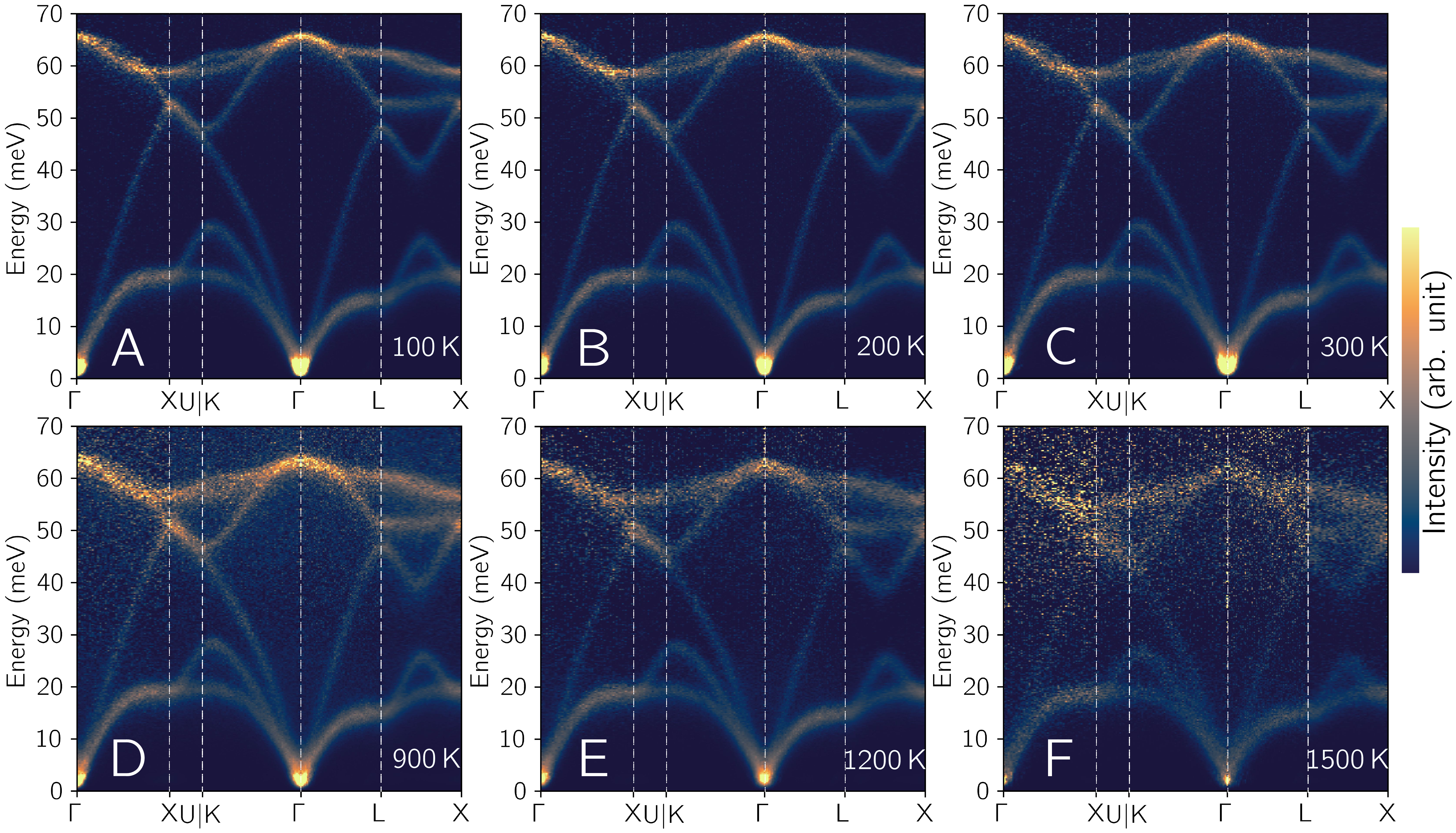}
\caption{Experimental phonon dispersions of silicon. Inelastic neutron scattering data of silicon were measured on the ARCS time-of-flight spectrometer at (A) 100\,K, (B) 200\,K, (C) 300\,K, (D) 900\,K, (E) 1200\,K, (F) 1500\,K. The 4-D phonon dynamical structure factor, \textit{S}(\textbf{\textit{q}},$\varepsilon$), were reduced, multiphonon subtracted, and ``folded'' into one irreducible wedge in the first Brillouin zone. Phonon dispersions are shown along high symmetry lines and through the zone (L--X).}
\label{fig:data}
\end{figure*}

\begin{figure*}
\includegraphics[width=\textwidth]{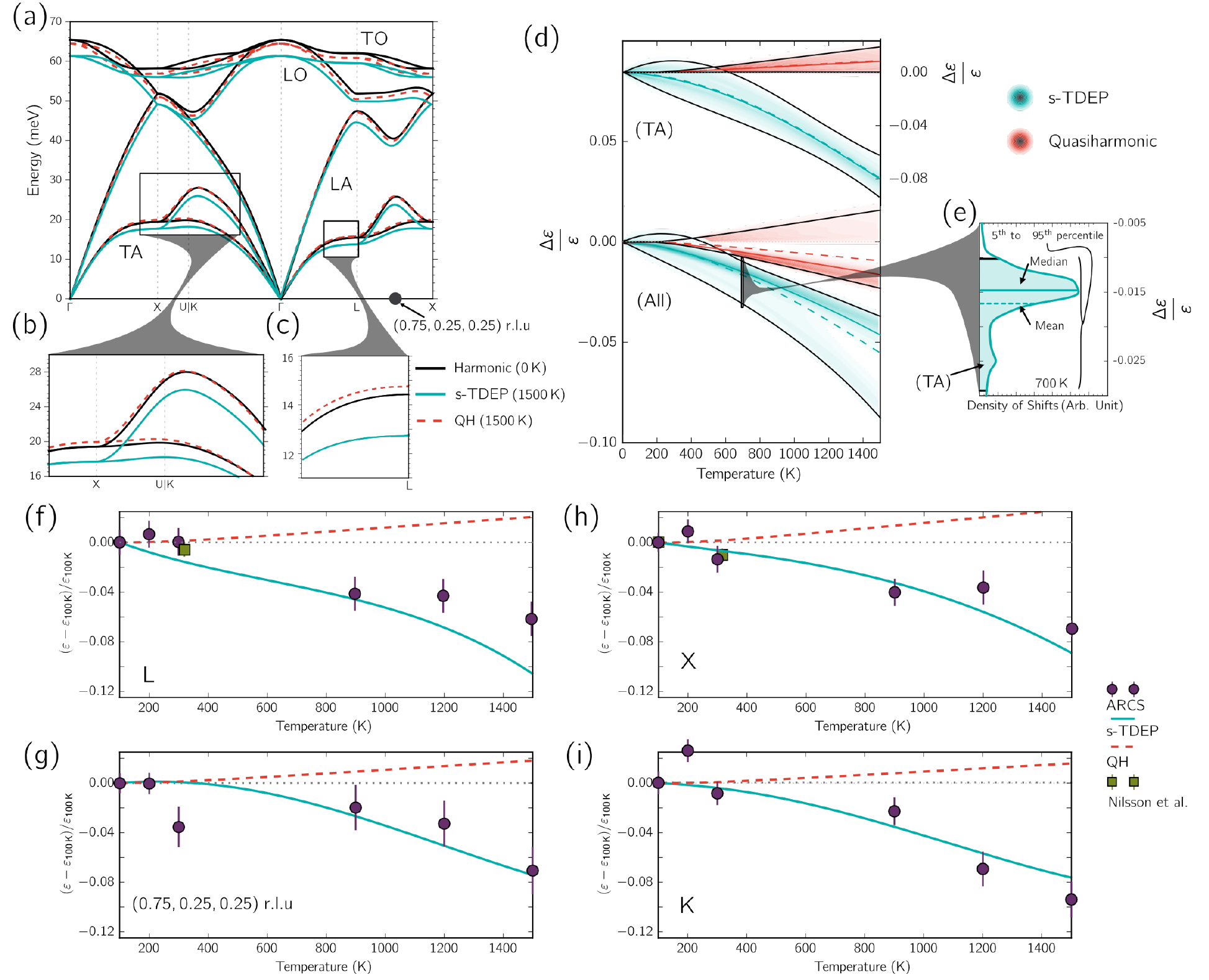}
\caption{Comparison between experimental, s-TDEP and quasiharmonic (QH) \textit{ab initio} calculations throughout the Brillouin zone. (A)--(C), Phonon dispersions of silicon from harmonic, s-TDEP, and QH \textit{ab initio} density functional theory calculations. The (0.75,0.25,0.25)-point is shown as a black circle marker for reference. (B),(C) Insets shows low-energy transverse acoustic modes. (D) Density of fractional phonon energy shifts with temperature. The densities from all branches (s-TDEP: teal, QH: red) and densities from just the low transverse modes are offset and scaled for clarity. (E) The density of shifts from s-TDEP at 700\,K. Notice the more negative peak consists of a majority of TA-modes. Temperature-dependent phonon shifts, $(\varepsilon-\varepsilon_{\rm{100\,K}}) / \varepsilon_{\rm{100\,K}}$, of the low-energy transverse modes at the (F) L, (G) X, (H) K, and (I) (0.75,0.25,0.25) r.l.u. points. Experimental fits of phonon centroids with standard (1\,$\sigma$)
error-bars from the present work are shown alongside calculated shifts and previously reported shifts \cite{Nilsson:1972hn}.}
\label{fig:fig2}
\end{figure*}

The cubic and quartic, and higher-order terms of the interatomic potential, cause the normal modes to interact and exchange energy.
This is pure anharmonicity, where the energy of a phonon is altered by the presence of other phonons irrespective of the volume of the solid.
Phonon anharmonicity is essential for thermal conductivity and other thermophysical properties.
Anharmonic effects increase with larger thermal atomic displacements.
Sometimes this causes a misperception that pure anharmonicity is important only at high temperatures, and quasiharmonic models may be valid at low and moderate temperatures owing to low phonon populations.
However, the leading-order terms of both quasiharmonicity and anharmonicity are linear in temperature \cite{maradudin1}, so if anharmonicity is important at high temperatures, it can have the same relative importance at low temperatures, too.
Furthermore, at low temperatures the ``zero-point'' energy gives atom displacements that allow a nuclear quantum effect to engage the high-energy phonon modes that are half occupied.

Finding the relative importances of quasiharmonicity and anharmonicity should be done by quantitative analysis, but to date the dominance of quasiharmonicity for silicon has been assumed in part because quasiharmonic models predict the thermal expansion with reasonable accuracy \cite{biernacki,fleszar,SiqingWei:2011tb,Liu:2014dr,Xu.PRB.1991s,baroni_revmod,rignanese}.
The quasiharmonic model predicts the anomalous negative thermal expansion of silicon from 10\,K to 125\,K and the low thermal expansion up to the melting temperature \cite{Middelmann:2015in,Okada:1984cz,Slack:1975js,Batchelder:1964ht,1972SSCom..10..159S}.
The positive thermal expansion coefficients observed at moderate and high temperatures are anomalous in their own right -- they are small compared to diamond and other materials with zincblende structures  \cite{Slack:1975js}.
Further validation of the quasiharmonic approximation was provided by measurements of the Raman mode and a few second-order Raman modes of silicon under pressure, which were accurately predicted by volume-dependent density functional theory (DFT) calculations at low temperature \cite{Hellman:2013di,Weinstein.PRB.1975}.
The negative Gr\"{u}neisen parameters of the low-energy transverse acoustic (TA) modes have received considerable attention and have been attributed to the ``open-ness'' of the diamond cubic structure \cite{1972SSCom..10..159S}, the stability of angular forces \cite{Xu.PRB.1991s}, or entropy in general \cite{Liu:2014dr}.
Nevertheless, the precise role of the TA modes in thermal expansion remains unclear \cite{SiqingWei:2011tb,Xu.PRB.1991s}.
With increasing temperature, phonons are excited in higher-energy phonon branches, and their positive Gr\"uneisen parameters are expected to cause the overall thermal expansion to change sign.
Today this quasiharmonic model is the workhorse for predicting thermal expansion.

``Non-trivial'' phonon shifts that were not accounted for by thermal expansion were reported in an earlier experimental paper on phonon dispersions in silicon up to 300\,K  \cite{Nilsson:1972hn}.
The importance of pure anharmonicity in temperature-dependent phonon shifts at moderate and high temperatures was also found in work based on molecular dynamics, many-body perturbation theory, and \textit{ab initio} calculations on silicon \cite{debernardi,1991PhRvB..43.4541N,1983PhRvB..28.1928B,Wang:1989tz,Menendez:1984kj,1982ApPhL..41.1016T,Kim:2015fxba,debernardi_99,Lang:1999ki}.
The uncertainty principle and quantum distributions of nuclear positions influence the exploration of atomic potential landscapes.
The zero-point motion was shown to be important, but does not by itself reproduce the correct thermal expansion coefficients \cite{Allen:2015dm,Herrero:2014jy}.
Temperature-dependent phonon shifts from pure phonon anharmonicity with zero-point energy could give a nuclear quantum effect that alters thermophysical properties.
A more detailed study of the temperature dependence of phonons in silicon is therefore appropriate because very few modes were previously assessed \cite{Nilsson:1972hn,1982ApPhL..41.1016T,Menendez:1984kj,Brockhouse}.

\begin{figure*}
\includegraphics[width=0.8\textwidth]{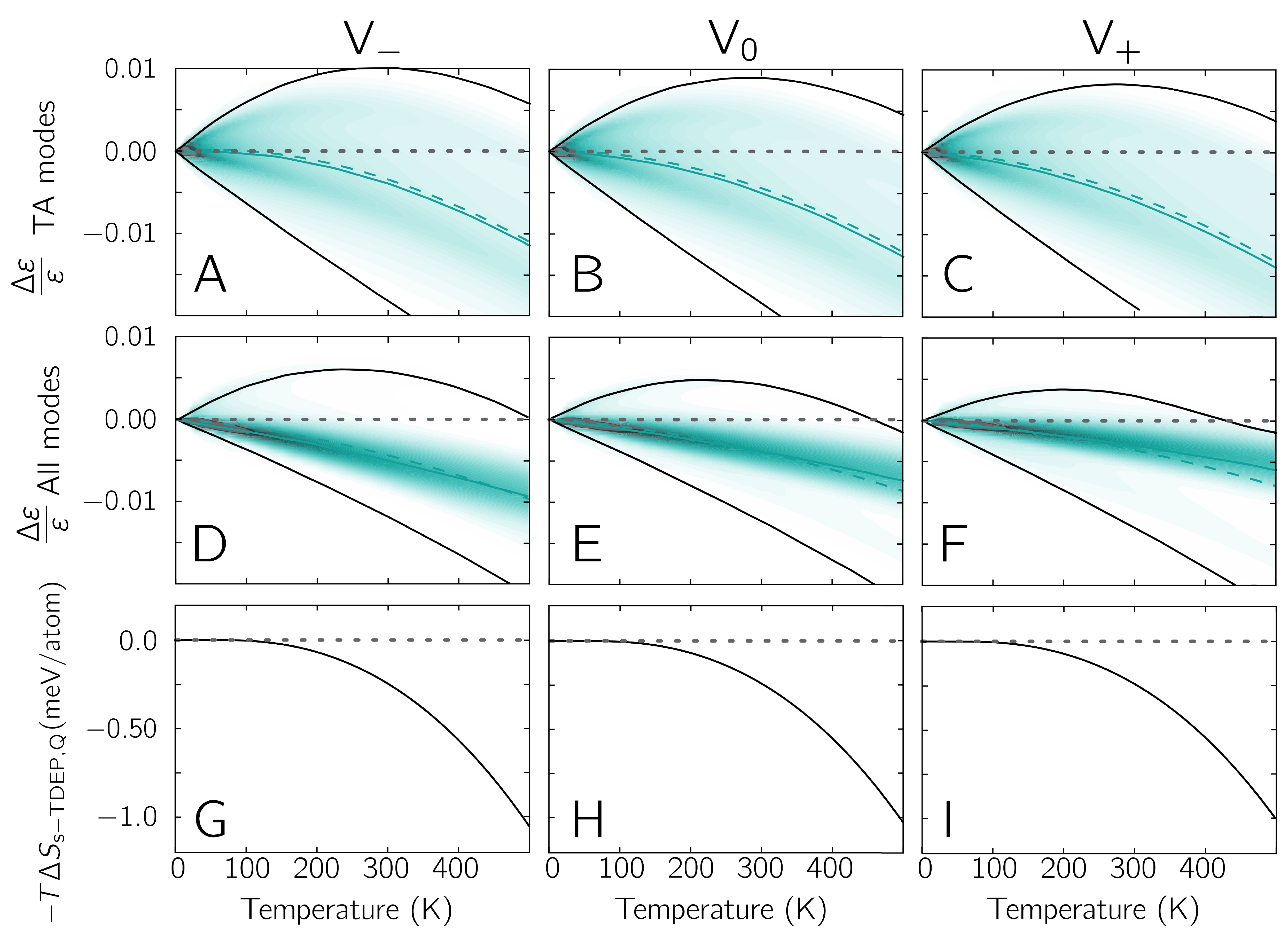}
\caption{Phonon shifts and entropy differences from constant volume \textit{ab initio} calculations. (A)--(F) Density of fractional shifts with temperature at constant volumes using the s-TDEP method. The mean (dashed color line), median (solid color line), and the 5$^{\textrm{th}}$ and 95$^{\textrm{th}}$ percentile (black solid lines) of the density are also shown. Calculations shown for: [(A),(D),(G)] 99\,\% of 0 K volume, [(B),(H),(E)] 0 K volume, and [(C),(F),(I)] 101\,\% of 0 K volume. Quasiharmonic predictions are the dashed zero--lines in (A)--(F). (G)--(I) Corresponding constant volume differences between the quasiharmonic (QH) and s-TDEP in free energies from vibrational entropy with temperature.}
\label{fig:fig4}
\end{figure*}
We report the first inelastic neutron scattering measurements of phonon dispersions of silicon above 300\,K along with fully anharmonic \textit{ab inito} calculations using the stochastically initialized temperature-dependent effective potential method (s-TDEP).
This stochastic method samples and fits the phonon potential landscape the same way a Born-Oppenheimer molecular dynamics potential energy surface is fitted to a model Hamiltonian \cite{Hellman:2013di}. 
This method can accurately describe highly anharmonic systems and includes higher order contributions of the lattice dynamic Hamiltonian, which intrinsically includes the phonon-phonon interactions as well as include the nuclear quantum effects \cite{Hellman:2013di,Klein1972,nina2017_tdep,errea,Wallace:1998vp}. 
These measurements are in conflict with the quasiharmonic theory, which predicts the wrong sign for phonon shifts with temperature.
We show that the crystal structure, quasiharmonicity, pure anharmonicity, and nuclear quantum effects all play important roles in the thermal expansion of silicon.
Methods for both the measurements and the calculations are described in the Materials and Methods and Supporting Information.

Figure\,\ref{fig:data} shows phonon dispersions as bright intensities.
The dispersions at low temperatures are in excellent agreement with previous work that used triple-axis spectrometers \cite{Nilsson:1972hn,Brockhouse}.
With increasing temperature, the majority of phonon modes, including the low-energy transverse acoustic modes, soften in proportion to their energy.
This self-similar behavior of phonon softening was reported previously \cite{Kim:2015fxba}.

Results from calculations by the s-TDEP method (with anharmonicity and thermal expansion) and conventional quasiharmonic \textit{ab initio} calculations (with no anharmonicity) are shown in Fig.\,\ref{fig:fig2}.
There are large discrepancies in the signs and magnitudes of phonon energy shifts between the two models.
Most interestingly, Fig.\,\ref{fig:fig2}\,(B),(C) show that the s-TDEP calculations predict a reduction in phonon energy, a thermal ``softening'', in the transverse modes (roughly $<$35 meV), whereas the quasiharmonic calculations predict an increase in phonon energy, ``stiffening'', at 1500 K (with negative Gr\"uneisen parameters as reported previously \cite{Xu.PRB.1991s,SiqingWei:2011tb,Liu:2014dr}).

We calculated the fractional shifts of energies, $\Delta \varepsilon / \varepsilon \left( T \right)$, for all phonon modes in the first Brillouin zone.
The energies of all phonons were calculated using a 50$\times$50$\times$50 grid of \textbf{\textit{q}}-points.
Figure\,\ref{fig:fig2}\,(D) compares the density of fractional phonon shifts from quasiharmonic and anharmonic (s-TDEP) calculations.
The density of fractional shifts, $\rho\left( \Delta \varepsilon/\varepsilon \right)$, is shown in Fig.\,\ref{fig:fig2}\,(E) from the s-TDEP method at 700\,K.
Compared to the quasiharmonic predictions for the TA modes (shown at top of Fig.\,\ref{fig:fig2}\,(D)), the anharmonic shifts are an order-of-magnitude larger, have opposite signs, and follow  opposite thermal trends.
Such large discrepancies allow for definitive experimental tests.

Individual phonon energies were obtained from constant-\textbf{\textit{q}} fits to the measured \textit{S}(\textbf{\textit{q}},$\varepsilon$), as shown in the Supporting Information.
Fig.\,\ref{fig:fig2}\,(F)--(I) show that the trends from the anharmonic s-TDEP calculations are in  far better agreement with experiment than the quasiharmonic trends.
Thermal trends for individual phonons at the L,X,K-points (Fig.\,\ref{fig:fig2}\,(F)--(H)) are presented
for their importance in the interpretation of quasiharmonic results \cite{SiqingWei:2011tb}.
Another example for a phonon mode located away from a high-symmetry line  is shown in Fig.\,\ref{fig:fig2}\,(I).

Additional s-TDEP calculations of densities of thermal shifts suggest why the quasiharmonic theory has been so apparently successful.
Calculations were performed for volumes that were 1\% larger and 1\% smaller than the 0\,K harmonic volume calculated for Fig.\,\ref{fig:fig2}\,(A), and the results are shown on the left and right sides of
Fig.\,\ref{fig:fig4} for the TA modes (top three panels) and all phonon modes (middle three panels).
For all three volumes, at low temperatures there is a wide spread in the thermal phonon shifts, both stiffening and softening.
At low temperatures, the average thermal shift from anharmonicity at a fixed volume is surprisingly nearly zero.
At fixed volume, the shifts of all quasiharmonic phonons are zero, of course, so the two methods agree on the average.
This is seen in Fig.\,\ref{fig:fig4}\,(A)--(C) for the TA modes and in Fig.\,\ref{fig:fig4}\,(D)--(F) for all modes.
Nevertheless, the average phonon energies of the TA modes from the s-TDEP method show an ordinary softening with increased volume and increased temperature, inconsistent with the negative Gr\"uneisen parameters from  quasiharmonic calculations.
At high temperatures, Fig.\,\ref{fig:fig4}\,(D)--(F) show that all the modes tend to soften at similar rates.
Differences in vibrational entropies from the s-TDEP and quasiharmonic methods were calculated using equations in the Supporting Information.
The difference in entropies $\Delta S$ from the quasiharmonic and anharmonic calculations was used to obtain the $-T \Delta S$ shown in Fig.\,\ref{fig:fig4}\,(G)--(I).
For all volumes, the differences are negligible up to 125\,K but increase at higher temperatures (Fig.\,\ref{fig:fig4}).

\begin{figure}[h!]\center
\includegraphics[width=0.5\textwidth]{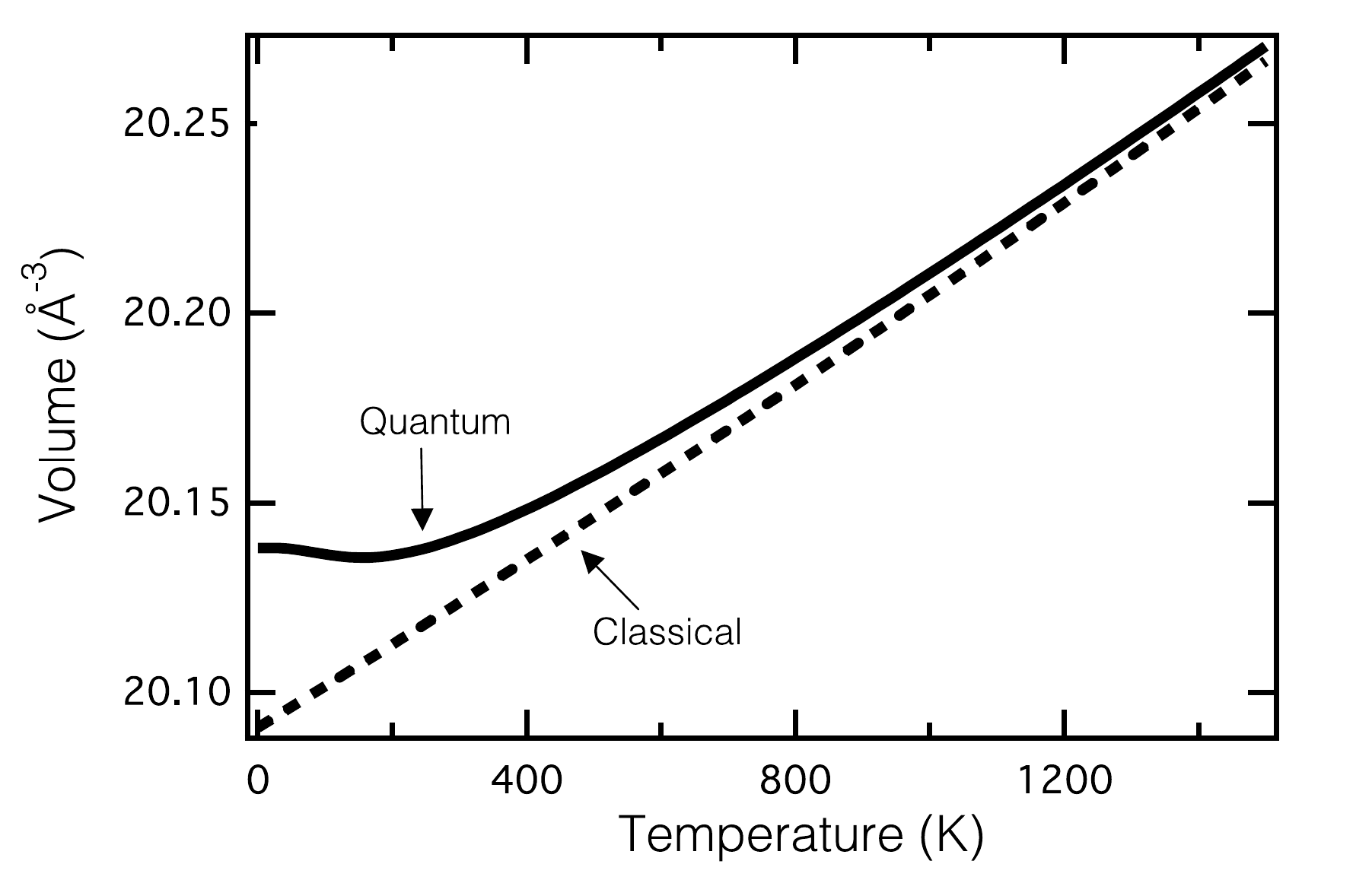}
\caption{Volume per atom as a function of temperature for silicon obtained from classical and quantum mechanical free energies.}
\label{fig:nqe}
\end{figure}

\begin{figure}[h!]\center
\includegraphics[width=0.5\textwidth]{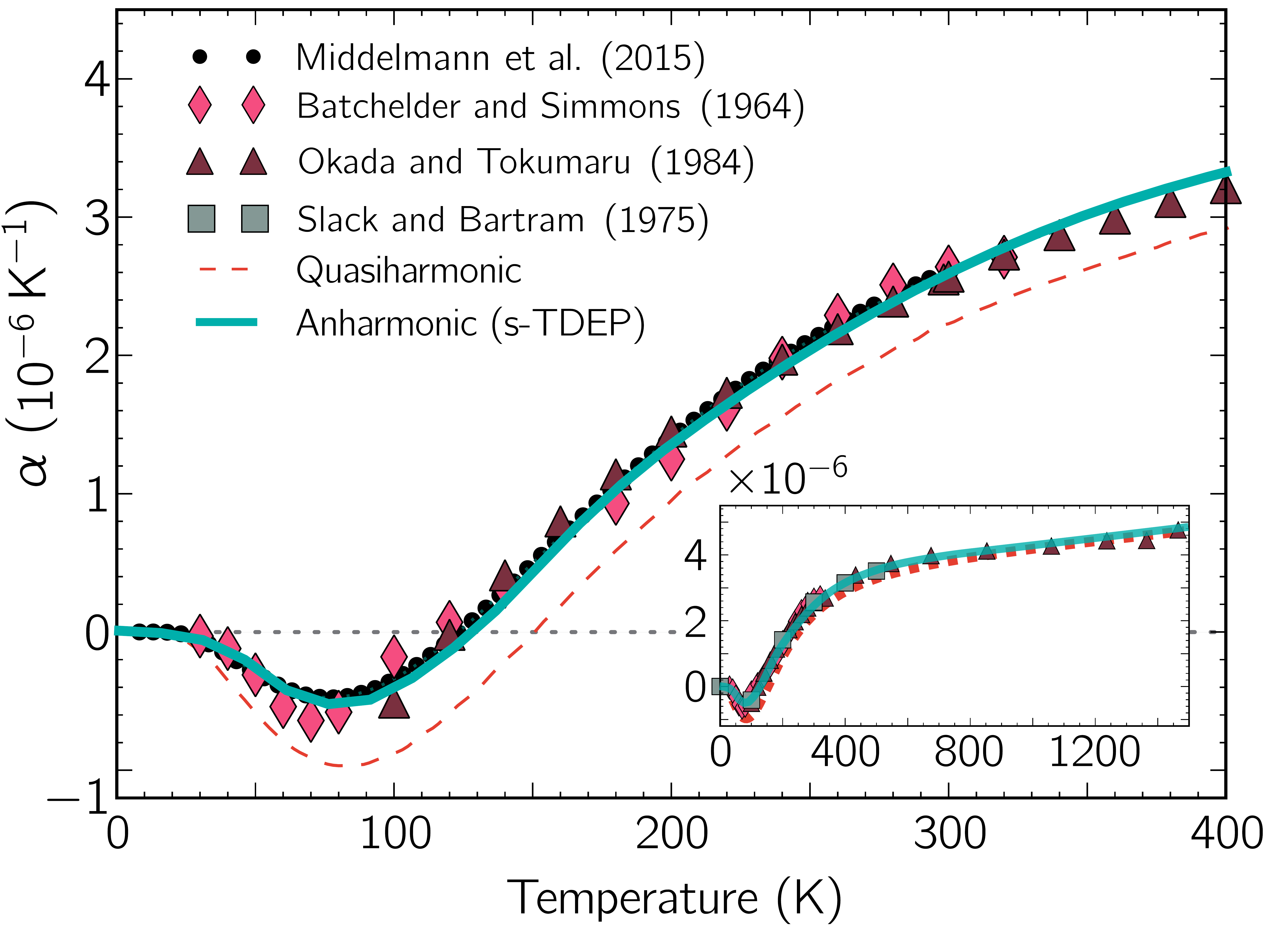}
\caption{Calculated and experimental coefficients of linear thermal expansion in silicon. Calculated coefficients are from minimized free energies using Supporting Information Eq.\,1 (s-TDEP: teal solid line, quasiharmonic (QH): red dashed line). Experimental values are shown as colored markers \cite{Middelmann:2015in,Okada:1984cz,Slack:1975js,Batchelder:1964ht}. Inset shows calculations and experimental values at higher temperatures.}
\label{fig:fig5}
\end{figure}
A quasiharmonic model with negative Gr\"uneisen parameters gives a physically incorrect explanation of thermal expansion, although some of its predictions of average properties are preserved by gross cancellations of errors.
As described in the Supporting Information, zero-point energy ($\hbar\omega_i /2$) in Eq.\,6 of the SI, proves essential for an anharmonic model to predict the negative thermal expansion of silicon (see Fig.\,\ref{fig:nqe}).
Nuclear quantum effects give nonzero anharmonic couplings between
all phonons, even modes of higher energy that are not excited thermally at low temperature.
These anharmonic couplings alter the self energies of the lower-energy phonons
that are excited at low temperatures, altering the
volume-dependence of the free energy.
Calculated coefficients of linear thermal expansion are in excellent agreement with experiments (Fig.\,\ref{fig:fig5}).   
Not only are quantum effects essential at lower temperatures, but differences persist up to melting temperatures.
Varying the zero-point motion from changes in nuclear mass allow for an interesting engineering
opportunity, too \cite{Herrero:2009gl,Herrero:2014jy,Herrero:1999wx,Noya:1997fo}.

Measurements of the phonon dispersions of single crystal silicon from 100 to 1500\,K showed thermal shifts that contradict the trends predicted by the widely accepted quasiharmonic model, even at low temperatures.
Pure phonon anharmonicity, i.e., phonon-phonon interactions, dominate the phonons in silicon from low to high temperatures, altering the effective interatomic potential and causing both positive and negative shifts of phonon energies.
At low temperatures the zero-point quantum occupancies of high-energy vibrational modes alter the energies of low-energy modes through anharmonic coupling. 
This nuclear quantum effect with anharmonicity and quasiharmonicity is the essential cause of the negative thermal expansion of silicon. 
The crystal structure, anharmonicity, and nuclear quantum effects all play important roles in the thermal expansion of silicon, and could be essential in other technologically important materials.

\nocite{Stone:2014kl,Abernathy:2012hf}
\nocite{Mantid,Arnold:2014iy}
\nocite{Squires:2012vi}
\nocite{Kim:2015fxba,Squires:2012vi}
\nocite{Kim:2015fxba}
\nocite{Nilsson:1972hn,debernardi,1982ApPhL..41.1016T}
\nocite{Ewings:2016ht,Squires:2012vi}
\nocite{Li:2015gfa,Hellman:2013di}
\nocite{Blochl:1994zz} 
\nocite{Kresse:1996kg,kresse_ab_1993,1994PhRvB..4914251K,Lee:1997ea}
\nocite{Armiento:2005kh,Mattsson:2009cf,Mattsson:2008gb}
\nocite{Sun:2016jp,Yao:2017gn}
\nocite{errea,Wallace:1998vp,Dove:1993wr}
\nocite{Hellman:2013di,hellman2}
\nocite{Hellman:2013di,errea}
\nocite{Kim:2015fxba}
\nocite{BrentFultz:2010ew}
\nocite{Abernathy:2012hf,Squires:2012vi}

\begin{acknowledgments}
The authors thank F.H. Saadi, A. Swaminathan, I. Papusha, and Y. Ding for assisting in sample preparation and discussions.
Research at Oak Ridge National Laboratory's SNS was sponsored by the Scientific User Facilities Division, BES, DOE.
This work used resources from NERSC, a DOE Office of Science User Facility supported by the Office of Science of the U.S. Department of Energy under Contract No. DE-AC02-05CH11231.
Support from the Swedish Research Council (VR) program 637-2013-7296 is also gratefully acknowledged. 
Supercomputer resources were provided by the Swedish National Infrastructure for Computing (SNIC).
This work was supported by the DOE Office of Science, BES, under contract DE-FG02-03ER46055.
\end{acknowledgments}


\bibliography{dskim89_si_alpha}

\clearpage

\subsection*{Supporting Information (SI)}
\subsection*{Inelastic Neutron Scattering}

Inelastic neutron scattering measurements were performed on a single crystal of silicon of  99.999\% purity that was highly-oriented ($<$2$^\circ$), purchased from Virginia Semiconductor, Inc.
The [110] oriented single crystal was further machined into a cylinder of 3.8\,cm in height, 2.54\,cm in outer diameter and a 1.59\,cm inner diameter to minimize multiple scattering.
The crystal was suspended in an aluminum holder and then mounted into a closed-cycle helium refrigerator for the 100 and 200 K measurements, and a similar holder made from niobium was mounted into a low-background electrical resistance vacuum furnace for measurements at 300, 900, 1200 and 1500\,K.
For all measurements the incident energy was 97.5\,meV, and an oscillating radial collimator was used to reduce background and multiple scattering \cite{Stone:2014kl,Abernathy:2012hf}.

The time-of-flight neutron data included multiple datasets from 200 rotations in increments of 0.5$^\circ$ about the vertical [110]-axis, reduced to create the 4-dimensional \textit{S}(\textbf{\textit{q}},$\varepsilon$)\,\cite{Mantid,Arnold:2014iy}.
A secondary data reduction process consisted of `folding' the entire \textit{S}(\textbf{\textit{q}},$\varepsilon$) data set into an irreducible wedge in the first Brillouin zone.
Non-linear offsets of the \textbf{\textit{q}}-grid were corrected by fitting typically 50 \textit{in situ} Bragg diffractions in an energy transfer range of $\Delta\varepsilon=\pm$ 4\,meV by a transformation to the positions of the theoretical diffraction peaks for a diamond cubic structure.
The multiphonon scattering was then subtracted, and the data `folded back' and corrected for the phonon creation thermal factor \cite{Squires:2012vi}.

The multiphonon scattering was determined with \textbf{\textit{q}}-dependence through the incoherent approximation and calculated from Eq.~\ref{eq:mph}\,\cite{Squires:2012vi},

\begin{equation}\label{eq:mph}
    S_{n>1} (\mathbf{{q}},\varepsilon) = \sum_{n=2}^{10} e^{-2W} \frac{(2W)^{n}}{n!}A_1 \circledast A_{n-1},
\end{equation} 

\noindent
where $2W$ is the well-known Debye-Waller factor calculated from the experimental temperature-dependent phonon density of states (DOS)\,\cite{Kim:2015fxba,Squires:2012vi}.
The single and $n$-phonon scattering spectrum are,

\begin{equation}
    A_1 = \frac{g(\varepsilon)}{\varepsilon} \langle n + 1 \rangle,
\end{equation}

\begin{equation}
    A_n = A_1 \circledast A_{n-1}.
\end{equation}

\begin{figure}[h!]\center
\includegraphics[width=0.5\textwidth]{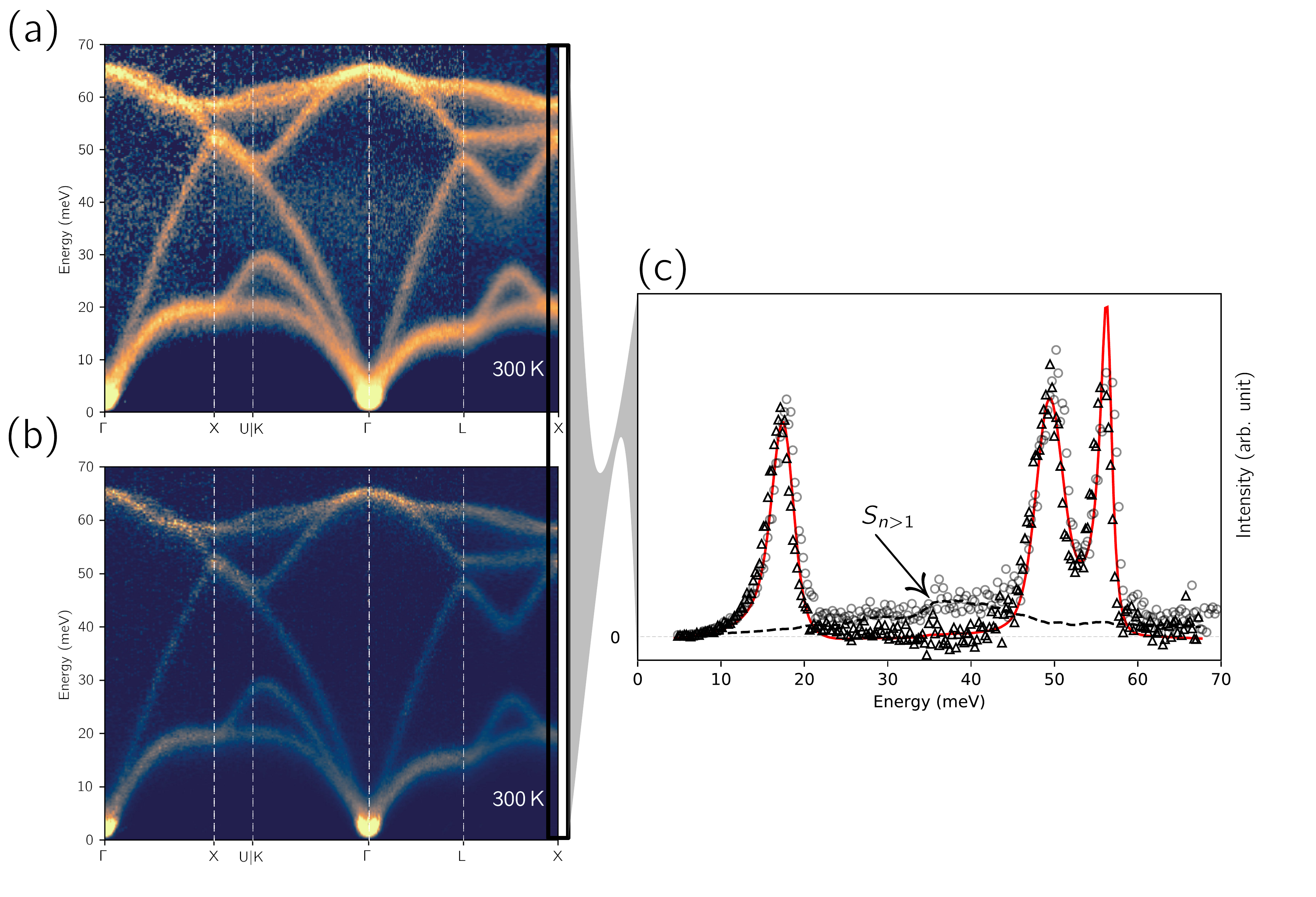}
	\caption{``Folded'' inelastic neutron scattering data without (a) and with (b) multiphonon subtracted \textit{S}(\textbf{\textit{q}},$\varepsilon$) at 300\,K. (c) Scattering intensity and fitted spectrum at the $X$-point. Fitted peaks are shown as the red solid line. Grey circles are without [(a)] and black triangles with [(b)] multiphonon scattering subtracted. Black dashed line shows subtracted multiphonon scattering intensity.}
\label{fig:mph}
\end{figure}

\noindent
The $g(\varepsilon)$ is the experimental phonon DOS\,\cite{Kim:2015fxba}, and $n$ is the Planck distribution. 
We find that even at temperatures $> 1000$\,K the contributions above the 5$^{\rm th}$ multiphonon spectrum ($S_5$) are negligible. 
A global scaling factor ($b*S_{n>1}$) was applied to the total multiphonon scattering function throughout the Brillouin zone after ``folding'' to correct for normalization. 
The multiphonon scattering accounted for most of the background intensity as seen clearly in Fig\,\ref{fig:mph}. 

The correct alignment of the data in reciprocal space and multiphonon subtraction produced \textit{S}(\textbf{\textit{q}},$\varepsilon$) of high statistical quality.
Thermal shifts of phonons reported previously, when available, were in good agreement\,\cite{Nilsson:1972hn,debernardi,1982ApPhL..41.1016T}.

\begin{figure}[h!]\center
\includegraphics[width=6.cm]{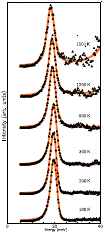}
	\caption{Constant \textbf{\textit{{q}}}-\textit{S}(\textbf{\textit{q}},$\varepsilon$) data at the X point for 100, 200, 300, 900, 1200, 1500\,K. Data are black markers and fits are in orange.}
\label{fig:fit}
\end{figure}

Energy spectra at specific \textbf{\textit{q}}-points were evaluated by integrating over 0.0025 \AA$^{-3}$.
Phonon centroids were then fitted using the Levenberg-Marquardt non-linear least square method for multiple skewed-Voigt functions.
The skewed-Voigt functions gave the best fits to the known asymmetric lineshape of the ARCS time-of-flight spectometer.
Examples of the scattered intensities at a constant-\textbf{\textit{q}}, with fits, are shown for the X-point in Fig.\,\ref{fig:fit}.

\begin{figure}[h!]\center
	\includegraphics[width=0.35\textwidth]{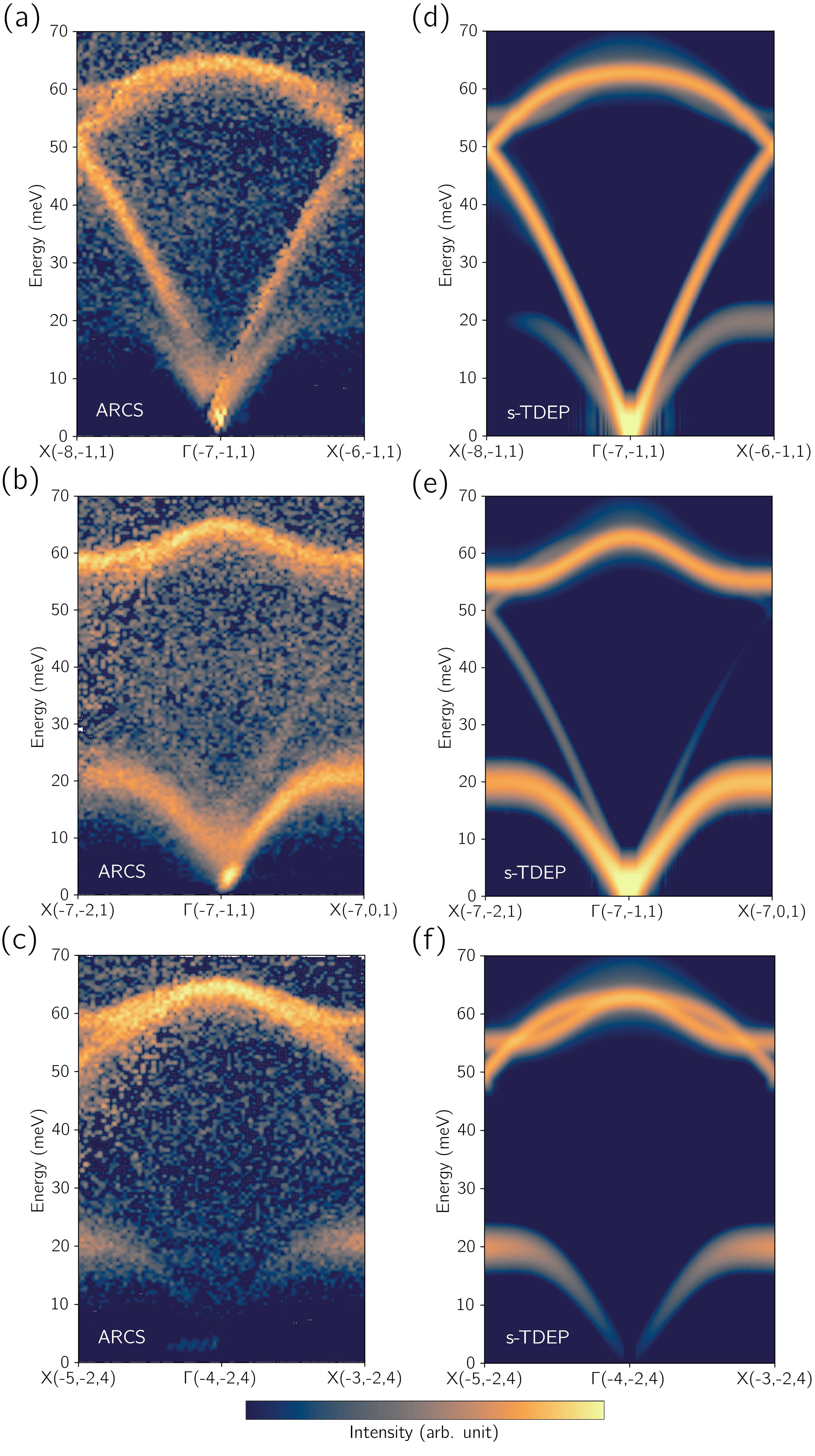}
	\caption{Inelastic neutron scattering [(a)--(c)] and first principles calculations [(d)--(f)] of \textit{S}(\textbf{\textit{q}},$\varepsilon$) at 300\,K along momenta \textbf{\textit{q}} (X--$\Gamma$--X) in different Brillouin zones. Calculated \textit{S}(\textbf{\textit{q}},$\varepsilon$) was corrected for instrument resolution and polarization effects to match the experiment conditions\,\cite{Abernathy:2012hf,Squires:2012vi}. }
\label{fig:slice}
\end{figure}

For comparison, a ``slice'' of ``unfolded'' 4-D \textit{S}(\textbf{\textit{q}},$\varepsilon$) along a momentum direction is shown in Fig.\,\ref{fig:slice}.
The data were processed using standard software and corrected for the phonon creation thermal factor\,\cite{Ewings:2016ht,Squires:2012vi}.
First principles calculations were performed using the s-TDEP method described below and elsewhere\,\cite{Li:2015gfa,Hellman:2013di}.
The experimental results are in good agreement with first principles calculations throughout reciprocal space.
There are benefits of assessing phonon intensities over multiple Brillouin zones, but these are not essential for a study of thermal expansion.

\subsection*{Ab-initio Calculations}

Ab initio DFT calculations were performed with the projector augmented wave \cite{Blochl:1994zz} formalism as implemented in VASP \cite{Kresse:1996kg,kresse_ab_1993,1994PhRvB..4914251K,Lee:1997ea}.
All calculations used a 5$\times$5$\times$5 supercell and a 500 eV plane wave energy cutoff.
The Brillouin zone integrations used a 3$\times$3$\times$3 $k$-point grid, and the exchange-correlation energy was calculated with the AM05 functional \cite{Armiento:2005kh,Mattsson:2009cf,Mattsson:2008gb}.
All calculations were converged to within 1\,meV/atom.
We found that calculations using other functionals as in\,\cite{Sun:2016jp,Yao:2017gn} gave similar phonon dispersion curves, and are expected to result in similar thermal trends.

Finite temperature phonon dispersions of silicon were calculated by fitting first-principles forces to a model Hamiltonian,
\begin{equation}\label{eq:test_ham}
\begin{split}
		H =\, &U_0 + \sum_i \frac{\mathbf{p}_i^2}{2m} + \frac{1}{2}\sum_{ij\alpha\beta}\Phi_{ij}^{\alpha\beta} u_i^\alpha u_j^\beta \\
			&+ \frac{1}{3!}\sum_{ijk\alpha\beta\gamma}\Phi_{ijk}^{\alpha\beta\gamma} u_i^\alpha u_j^\beta u_k^\gamma.
\end{split}
\end{equation}
The forces on atoms were generated using DFT with various configurations of displaced atoms by a stochastic sampling of a canonical ensemble, with Cartesian displacements ($u_i^\alpha$) normally distributed around the mean thermal displacement using
\begin{equation}\label{eq:1}
	u_i^\alpha = \sum_k \frac{\epsilon_k^{i\alpha} c_k}{\sqrt{m_i}} \sqrt{-2\ln\xi_1} \sin( 2\pi\xi_2 ).
\end{equation}
The thermal factor, $c_k$, is based on thermal amplitudes of normal mode $k$, with eigenvector $\epsilon_k$ and frequency $\omega_k$ \cite{errea,Wallace:1998vp,Dove:1993wr}
\begin{equation}\label{eq:2}
	c_k = \sqrt{\frac{\hbar( 2n_k + 1)}{2\omega_k}},
\end{equation}
and $\xi_1$ and $\xi_2$ are stochastically sampled numbers between 0 and 1.
The phonon distribution follows the Planck distribution, $n_k=(e^{\beta \hbar \omega_k} - 1)^{-1}$, where the nuclear quantum effect can be turned off by taking the high-temperature limit of Eq.\,\ref{eq:2}.
The fitting to the model Hamiltonian used the temperature-dependent effective potential method (TDEP) \cite{Hellman:2013di,hellman2}.
With thermal displacements from Eq.\,\ref{eq:1} and Eq.\,\ref{eq:2}, we refer to our temperature-dependent calculations as the stochastically-initialized temperature-dependent effective potential method (s-TDEP).

This method circumvents the issue of expensive computational resources required of \textit{ab initio} molecular-dynamics (AIMD), replacing AIMD with a Monte Carlo sampling of atomic positions and momentum near equilibrium positions \cite{Hellman:2013di,errea}.
The quasiharmonic model was calculated as described previously \cite{Kim:2015fxba}.

\subsection*{Phonon Shifts}
The temperature dependent renormalized phonon frequencies were calculated from the phonon self-energy.
The phonon self-energy part, corrections to phonon energies from many-body interactions, is comprised of real and imaginary contributions, 

\begin{equation}\label{eq:self}
\Sigma_\lambda = \Delta_\lambda + i \Gamma_\lambda.
\end{equation}
Phonon scattering rates, and phonon lifetimes, are related to the imaginary part of the self-energy ($\frac{1}{\tau_{\lambda}}=2 \Gamma_{\lambda}$) for mode $\lambda$ evaluated at the harmonic frequency. 
The imaginary part of the self-energy from many-body perturbation theory is 

\begin{equation}
\begin{split}
\Gamma_{\lambda}(\Omega)  & =  \frac{\hbar\pi}{16}
\sum_{\lambda'\lambda''}
\left|
\Phi_{\lambda\lambda'\lambda''}
\right|^2 
\\& \times \big{\{}(n_{\lambda'}+n_{\lambda''}+1)
\delta(\Omega-\omega_{\lambda'}-\omega_{\lambda''})+
(n_{\lambda'}-n_{\lambda''})
\\& \times \left[
\delta(\Omega-\omega_{\lambda'}+\omega_{\lambda''}) -
\delta(\Omega+\omega_{\lambda'}-\omega_{\lambda''})
\right]
\big{\}}.
\end{split}
\end{equation}
The $\Omega(=E/\hbar)$ is the probing energy and the delta functions conserve energy and momentum and sum over all possible three-phonon interactions between modes.
The $\Phi_{\lambda\lambda'\lambda''}$ is the three-phonon matrix element, the Fourier transform of the third-order component of the interatomic potential (Eq.\,\ref{eq:test_ham}).
The real part of the self-energy is calculated through a Kramers-Kronig transformation of the imaginary part 
\begin{equation}
    \Delta(\Omega) = \frac{1}{\pi} \int d\omega \frac{\Gamma(\omega)}{\omega-\Omega}.
\end{equation}      

The probing frequency, $\hbar\Omega$, is directly comparable to the renormalized frequencies from the harmonic energies with perturbative shifts calculated from the cubic term of the model Hamiltonian \cite{Wallace:1998vp}, 
\begin{equation}
	\hbar \Omega = \hbar(\omega_{k} + \Delta_{k}).
\end{equation}
Essentially, the effective potential method fits the stochastically sampled phonon potential to capture all even terms in the phonon potential as the ``harmonic'' frequencies ($\omega$), and the odd-term contributions add shifts of the frequencies perturbatively ($\Delta$) \cite{Klein1972,nina2017_tdep}.
This method includes higher-order phonon-phonon interactions by renormalizing terms in the model Hamiltonian.

\subsubsection*{Thermodynamic Calculations}
Temperature dependent coefficients of linear thermal expansion in silicon were calculated through the minimization of the free energy,

\begin{equation} \label{eq:free}
\begin{split}
{F}_{}({T},{V}) = & \, {E}({T,V}) + \sum_{\mathbf{q},k} \Big( \frac{\hbar\omega_k(\mathbf{q},V,T)}{2} \\ &+ {k}_{\mathrm B} {T}
 \ln (1- e^{-\hbar\omega_k(\mathbf{q},T,V) /{k}_{\mathrm B} {T}}) \Big),
\end{split}
\end{equation}
from quasiharmonic calculations, and from s-TDEP (main text Fig.\,5).
The quasiharmonic model assumes the only temperature dependence of the entropy is from the volume expansion $\varepsilon_k(V\{T\})$ and the Planck distribution ($n_k$), whereas the anharmonic s-TDEP method minimizes the free energy for temperature and volume simultaneously.
The vibrational entropy from all phonon modes, $\sum_k$, was calculated as \cite{BrentFultz:2010ew},
\begin{equation} \label{eq:entropy}
{S}_{\rm vib}({T}) = 3 {k}_\mathrm{B} \sum_k \big[ (n_k+1) \ln ( n_k+1) - n_k \ln( n_k)     \big].
\end{equation}

\subsection*{Possible Contributions to the Thermal Expansion}
As stated in the main text, a simple physical model for the anomalous thermal expansion of silicon is unlikely because different effects contribute to the thermal expansion.
In particular, the anharmonicity and nuclear quantum effects are difficult to formulate as a simple 3D model.
The thermal expansion of Si can be simulated properly with methods based on \textit{ab initio} calculations that includes all these factors, but this seems unsatisfying for a ``physical'' understanding.
A number of possible contributions and models are presented here, but any single model is insufficient by itself.

\subsubsection*{Negative Gr\"uneisen Parameters as Fitting Parameters}

\begin{figure}\center
\includegraphics[width=0.45\textwidth]{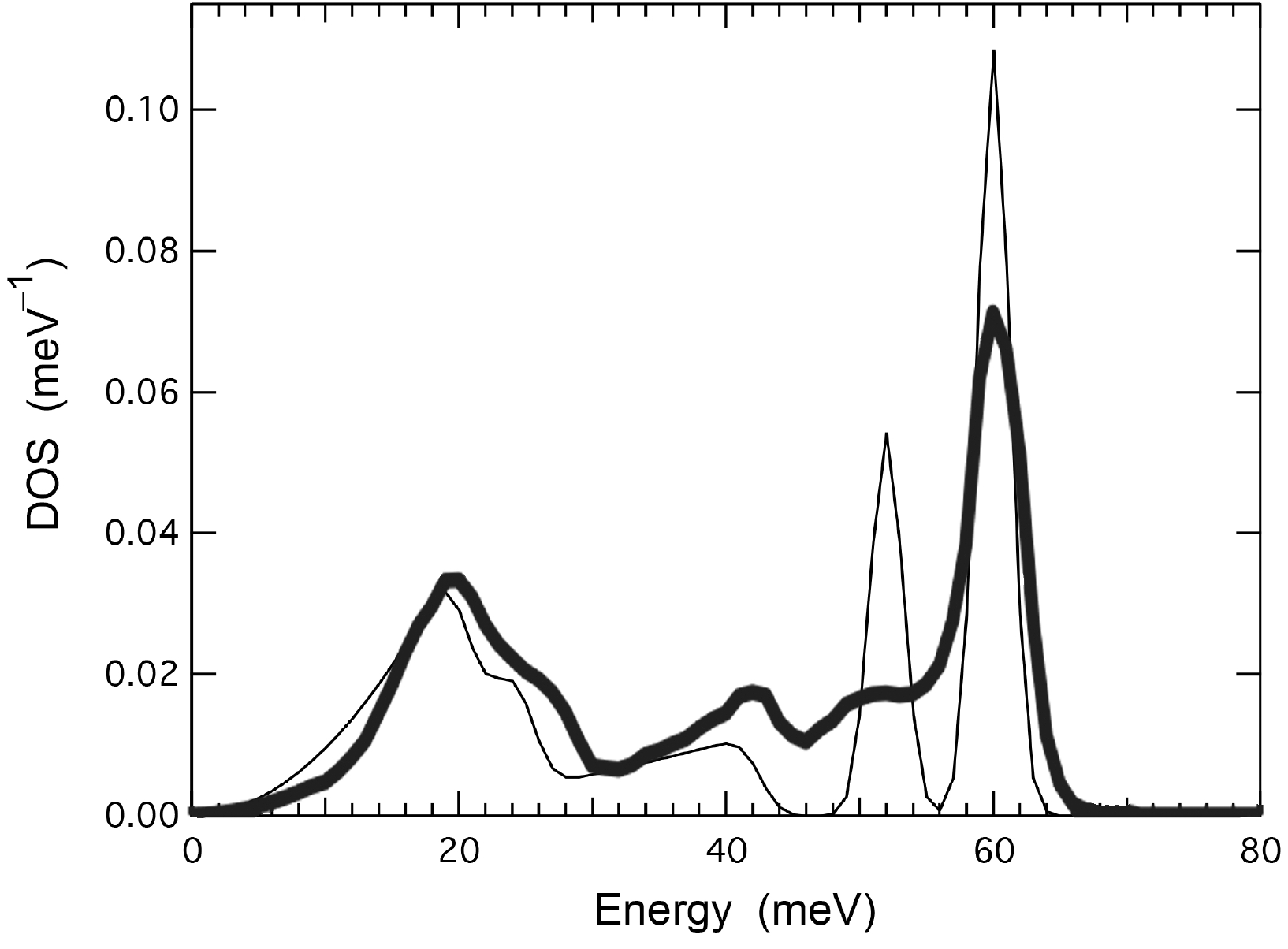}
\caption{Phonon DOS of Si from (thick line) experimental measurement at 100\,K \cite{Kim:2015fxba}, and (thin line) approximated with Debye and Einstein models.}
\label{SupplSiDOS}
\end{figure}

If individual Gr\"uneisen parameters are assigned to different parts of the phonon
DOS, it is easy to make a model that predicts negative thermal expansion at low temperatures.
An approximation for silicon is shown in Fig.\,\ref{SupplSiDOS},
together with the experimental phonon DOS reported previously \cite{Kim:2015fxba}.
The six phonon branches were modeled as follows:
acoustic branches were approximated by Debye models
with cutoff energies of 20, 25, 42\,meV, and
optical branches were approximated as Einstein modes
with energies of 52, 60, 60 meV.
These curves were convoluted with a Gaussian function
of standard deviation $\sqrt{3} \,$meV, summed, and are
compared to the experimental phonon DOS of Si \cite{Kim:2015fxba} in
Fig.\,\ref{SupplSiDOS}.

\begin{figure}\center
\includegraphics[width=0.4\textwidth]{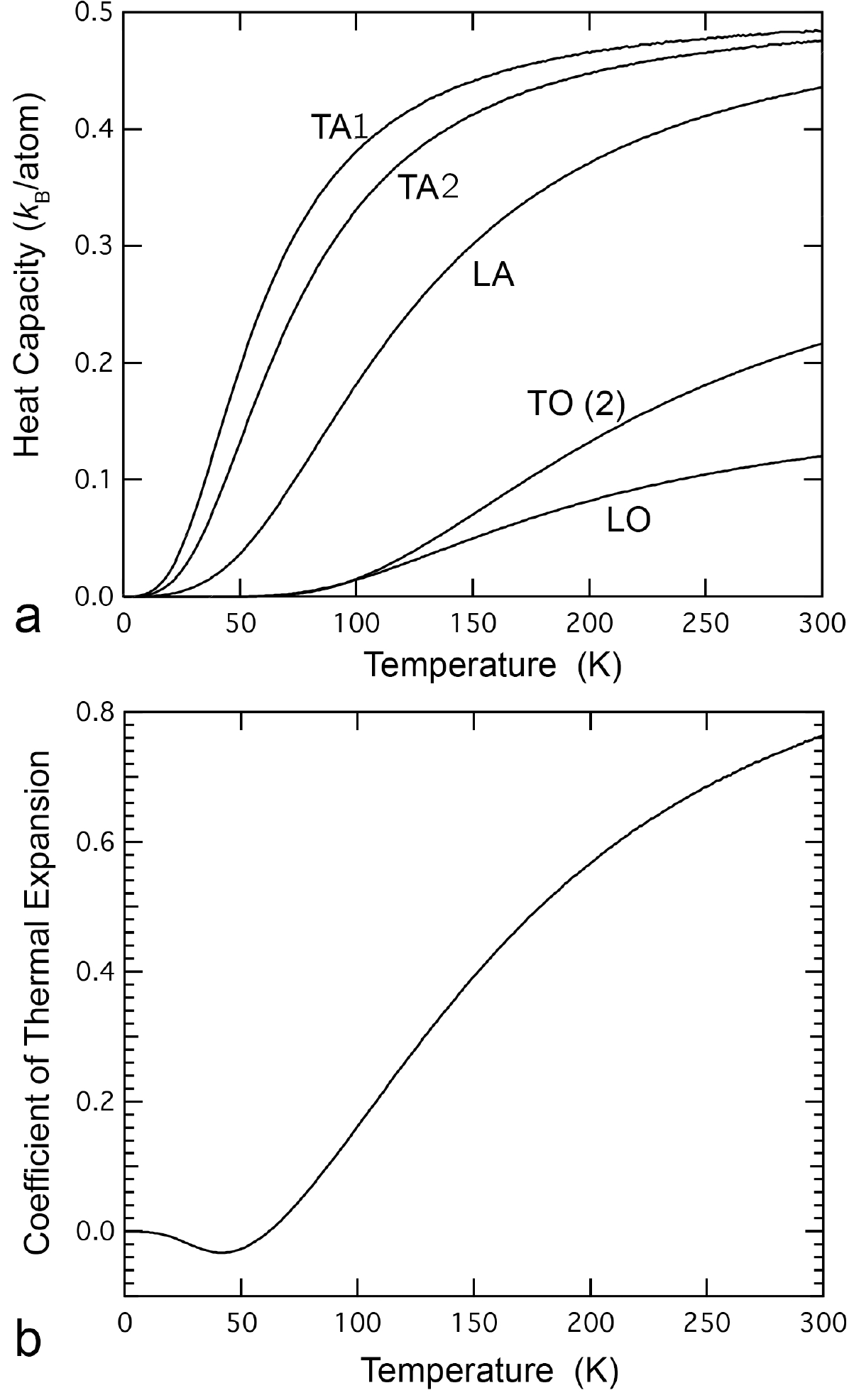}
\caption{({a}) Heat capacities from phonons approximated by Debye and Einstein models, using the six branches of Fig.\,\ref{SupplSiDOS}.
({b}) Coefficient of thermal expansion, assuming all Gr\"uneisen parameters were +1 except for TA modes set as --1.}
\label{Suppl_HeatCapTherExp}
\end{figure}

The heat capacities of these six functions were calculated as shown in Fig.\,\ref{Suppl_HeatCapTherExp}.(a).
For simplicity, a Gr\"uneisen parameter of --1 was assigned to the lowest-energy TA modes,
and a Gr\"uneisen parameter of +1 was assigned to the other five phonon branches.
The thermal expansion as a function of temperature, shown in Fig.\,\ref{Suppl_HeatCapTherExp}.(b),
has a shape that follows the heat capacity curves times their Gr\"uneisen parameters.
At low temperatures, the negative contribution from the TA1 modes overcomes the
positive contribution from the TA2 modes, but the thermal expansion changes sign when the LA modes
are sufficiently occupied.
With six Gr\"uneisen parameters, there are many ways to optimize the thermal expansion as a function
of temperature, and the depth and breadth of the minimum can be tuned by appropriate parameter selection.
We did not explore this further because the main text shows that this approach is physically incorrect.

\subsubsection*{Simple Springs and Angular Bonds}

\begin{figure}\center
\includegraphics[width=0.4\textwidth]{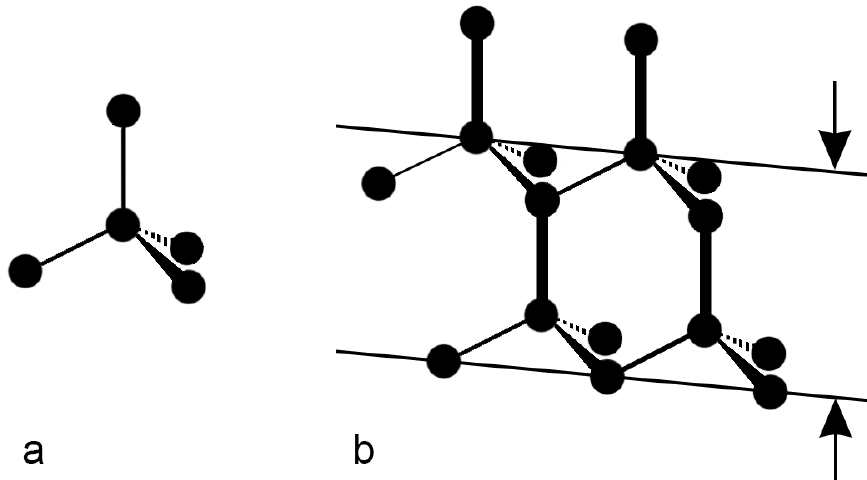}
\caption{\textbf{a}, Tetrahedral coordination around a central Si atom. \textbf{b}, four interconnected tetrahedra of the diamond cubic structure. Thick vertical lines are along a  $[ 111 ]$ direction.}
\label{Suppl_tetrahedra2}
\end{figure}

The simplest model of harmonic interatomic forces is useful for illustrating a geometrical source of phonon anharmonicity.
Fig.\,\ref{Suppl_tetrahedra2}.(a) shows a tetrahedron with a Si atom surrounded by its nearest neighbors.
The four bonds are assumed to be harmonic springs,
and  it can be initially assumed that the neighbors remain fixed in position.
As shown in Fig.\,\ref{Suppl_tetrahedra2}.(a), the springs are  relaxed, with no elastic energy.
If the central Si atom is displaced vertically, the amount of elastic energy stored in the spring to the neighbor above is the same for positive and negative displacements of equal magnitude.
This symmetry does not hold for the lower three springs.
Upwards displacements are more along the directions of the springs,
and generate more elastic energy than downwards displacements.
The elastic energy is straightforward to calculate for harmonic springs in a tetrahedral coordination
with angles of 109.5$^{\circ}$ and a nearest-neighbor separation of $a$.
For vertical displacements, $x$, a numerical fit to the elastic energy in all four springs gives
\begin{eqnarray}
E_{\rm el}(x) \propto \left( \frac{x}{a} \right) ^2 + 0.666 \left( \frac{x}{a} \right)^3 \; , \label{TetrahedralAnharmonicity}
\end{eqnarray}
so negative $x$ (downwards) displacements are more favorable energetically.
The lengthening of the vertical bond in Fig.\,\ref{Suppl_tetrahedra2}.(a)
gives positive thermal expansion, and it is likely important at high temperatures
when numerous short wavelength phonons disrupt the cooperative displacements between adjacent
tetrahedra.

For long wavelength phonons, however, displacements along the $[ 111 ]$ direction can provide for
negative thermal expansion.
Figure\,\ref{Suppl_tetrahedra2}.(a) helps to illustrate a phonon mode where the vertical Si pairs along $[ 111 ]$ maintain a fixed
separation, and vibrate as a unit along the $[ 111 ]$ direction.
For the case shown in Fig.\,\ref{Suppl_tetrahedra2}.(b), the cubic anharmonicity of
Eq.\,\ref{TetrahedralAnharmonicity} will cause a
decrease in separations between  planes of atoms, illustrated by the arrows.
For a 1\% mean-squared displacement, Eq.\,\ref{TetrahedralAnharmonicity} predicts that
the cubic term is approximately 1\% of the magnitude of the quadratic.
We might expect the lattice parameter to decrease by approximately
one part in 10$^{-4}$ if such modes dominate.
The complexity of accounting for all different modes and their thermal occupancies
makes further analysis impractical, however.
These examples show
\begin{itemize}
\item
When atom displacements are not along the directions of the springs,
phonon modes can be anharmonic even when the springs are harmonic.

\item
These geometrically-induced anharmonicites can change sign with
the wavevector of the phonon mode.

\end{itemize}

The influence of anharmonicity may also be expected because the geometrical structure of silicon does not have inversion symmetry at each atom.
This allows cubic phonon-phonon interactions in first order \cite{maradudin1,feldman}, making vibrational modes more free to exchange energy.

\begin{figure}\center
\includegraphics[width=0.4\textwidth]{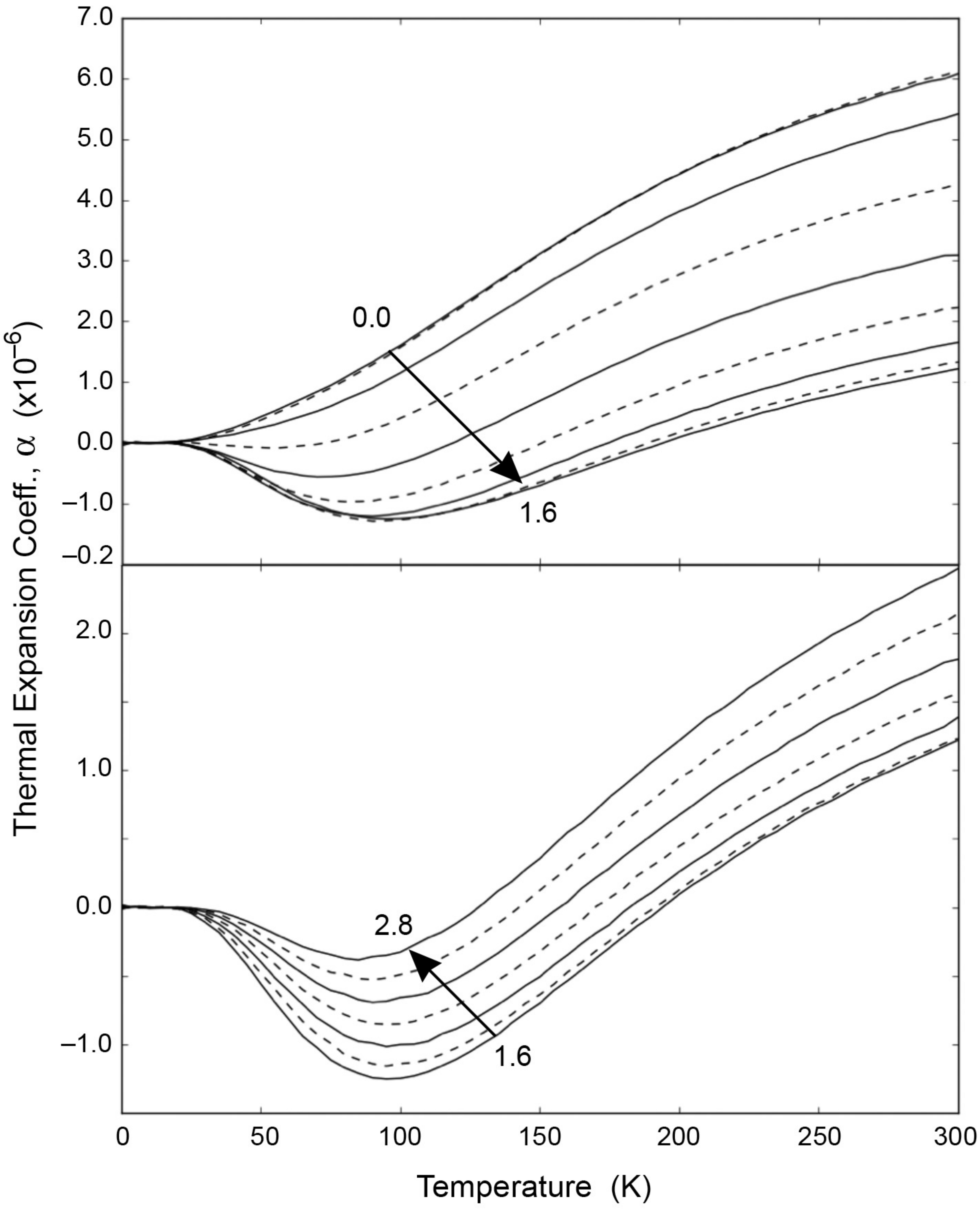}
\caption{Trends for silicons thermal expansion coefficient vs. temperature plotted for scalings of the ratio of transverse to longitudinal force constants between 0 and 2.8}
\label{Suppl_ratios}
\end{figure}
\begin{figure}\center
\includegraphics[width=0.4\textwidth]{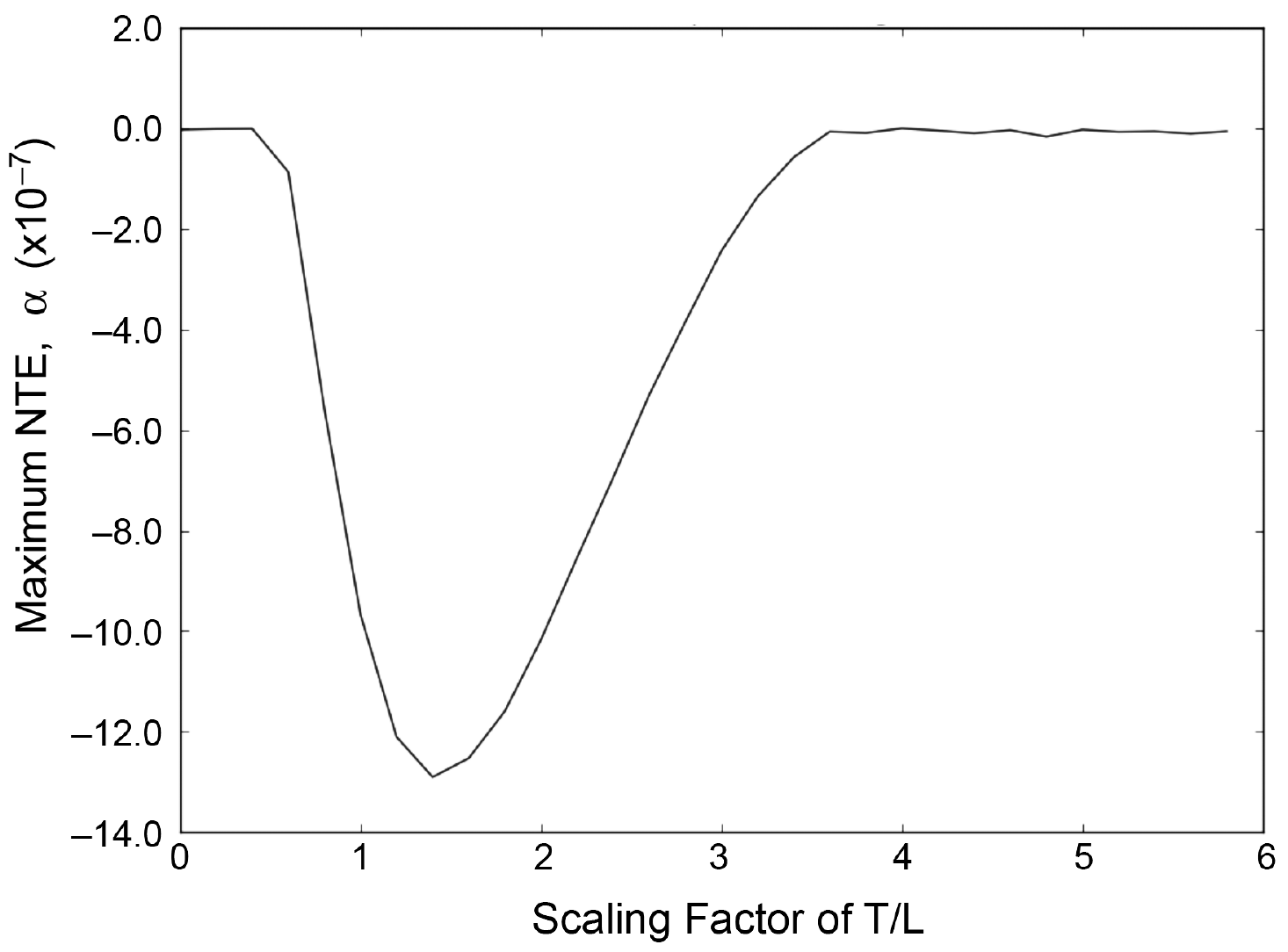}
\caption{Maximum negative thermal expansion coefficients taken from each trend in Fig. S5 plotted against the scaling factor, $k$, applied to the ratios of transverse to longitudinal force constants.}
\label{Suppl_ratios2}
\end{figure}
\subsubsection*{Transverse Bonds}\label{sec:transv}

The diamond cubic structure is not stable under longitudinal forces alone, and transverse forces are required to prevent its collapse into a denser structure.
Xu, et al., argue that the negative thermal expansion of silicon depends on the relative strengths of the first-nearest-neighbor bond-bending and bond-stretching forces \cite{Xu.PRB.1991s}.
For two specific phonon modes (TA(\textbf{\textit{q}}=X) and TA(\textbf{\textit{q}}=L)), they develop a mechanical model that predicts negative Gr\"uneisen parameters when both central forces and non-central forces are of comparable strength.
They note that the relative strength of the non-central forces plays a major role in setting the thermal expansion of silicon \cite{Xu.PRB.1991s}.

We decomposed pairwise interactions between silicon atoms, quantified as force constant tensors, into components that are transverse and longitudinal to the relevant bond.
Interestingly, we found that with a model quasiharmonic system there is some optimal scaling of the transverse to longitudinal force constants that results in maximal negative thermal expansion in Si.
(Although the simple modes described previously in Section B have no transverse forces but have negative Gr\"uneisen parameters, there are many other modes that contribute to the thermal expansion.)
To do this, we began with force constant tensors that describe pairwise interactions between the atom at each of the two distinct symmetry positions in silicon and its closest 123 neighbors as a function of volume at 0\,K.
For each pairwise interaction between each of the atoms at distinct symmetry positions and its neighbors, we decomposed the force constant tensor into components transverse and longitudinal to the bond between the pair of atoms.
We scaled the ratio of transverse to longitudinal force constants by a constant, $k$, while holding fixed the norm of the force constant tensor.
We then calculated the the thermal expansion in silicon for values of k between 0 and 2.8.
In Fig.\,\ref{Suppl_ratios}, we show that increasing the ratios of transverse to longitudinal force constants in the system for values of $k$ between 0 and about 1.6 increases the amount of negative thermal expansion, and that increasing $k$ beyond 1.6 decreases negative thermal expansion.
In Fig.\,\ref{Suppl_ratios2}, we illustrate the dependence of the negative thermal expansion on the ratios of transverse to longitudinal force constants by plotting the minimum value of thermal expansion exhibited by the system (one metric for quantifying the degree of NTE) against the scaling constant $k$.
Although the quasiharmonic approximation should not be used to predict how negative Gr\"uneisen parameters give the negative thermal expansion of silicon, the ratio of forces should influence on the thermal expansion of diamond cubic structures.

\subsubsection*{Quantum and Zero-point Effects}

Models with transverse bonds and simple springs, like many other previous models, can be understood with classical mechanics.
There is evidence that this is inadequate \cite{Allen:2015dm,Herrero:2014jy}.
For lattice dynamics, the difference between quantum and classical particles are evident in the governing distributions.
Classical molecular dynamics or even Born-Oppenheimer \textit{ab initio} molecular dynamics are still classical depictions of nuclear dynamics, and do not provide the correct quantum distributions.
There have been great advances in overcoming this by utilizing path-integral methods to include nuclear quantum effects including zero-point motion\,\cite{Herrero:2014jy}, but a full \textit{ab intio} path-integral molecular dynamics of solid materials is computationally expensive.
We have addressed these limitations through our stochastic method as described above which includes the zero-point energy ($\hbar\omega_x /2$) in Eq.\,\ref{eq:2}, giving nuclear quantum effects with the anharmonicity.

Using Eq.\,\ref{eq:entropy} for classical or quantum distributions in Eq.\,\ref{eq:free} for the free energy give major differences in thermal expansion in silicon, as shown in Fig.\,\ref{fig:nqe}.
Even at lower temperatures, the zero-point energy brings importance to all the phonon modes.
Not only are quantum effects essential at lower temperatures, but differences persist up to melting temperatures.
Varying the zero-point motion from changes in nuclear mass allow for an interesting engineering opportunity, too\,\cite{Herrero:2009gl,Herrero:2014jy,Herrero:1999wx,Noya:1997fo} 

In general, all of the models explained above are effective for a pedagogical thought exercise for a ``physical interpretation'' of the negative thermal expansion of silicon, but no single simple model is able to capture the full behavior.
A simple model has not yet been provided, as there is none.

\end{document}